%% file: aa25839-15.tex
\documentclass[online]{aa}
\usepackage[varg]{txfonts}
\RequirePackage{hyperref}
\hypersetup{
	colorlinks = true,
	linkcolor = blue,
	citecolor = blue,
	filecolor = blue,
	urlcolor = blue,
	pdfborder = {0 0 0}
}
\newcommand{\doiurl}[1]{\href{https://doi.org/#1}{#1}}
\newcounter{IonCS}
\DeclareRobustCommand{\ion}[2]{\setcounter{IonCS}{#2}\textup{#1\,\large{\scshape{\roman{IonCS}}}}}
\graphicspath{{./}{../Bilder/}{../Bilder/profiles/}{../Bilder/core_combined/}}
\newcommand{\panel}[1]{\textit{#1~panel}:}

\newcommand{\sect}[1]{Sect.\,\ref{S:#1}}

\newcommand{\fig}[1]{Fig.\,\ref{F:#1}}
\newcommand{\figs}[2]{Figs.\,\ref{F:#1} and \ref{F:#2}}
\newcommand{\firstfig}[1]{\fig{#1}}

\newcommand{\eqn}[1]{Eq.\,\eqnr{#1}}

\newcommand{\eqnr}[1]{(\ref{E:#1})}

\newcommand{\graphflex}[4][figure]{\begin{#1}#2\caption{#4}\label{F:#3}\end{#1}}

\newcommand{\graph}[3]{\graphflex{\resizebox{\hsize}{!}{\includegraphics{#1.pdf}}}{#2}{#3}}
\newcommand{\graphwidthflex}[6][figure*]{\graphflex[#1]{#5\includegraphics[width=#4]{#2.pdf}}{#3}{#6}}
\newcommand{\graphwidth}[4][15cm]{\graphwidthflex{#2}{#3}{#1}{\centering}{#4}}

\newcommand{\graphreduced}[4][-6pt]{\graphflex{\resizebox{\hsize}{!}{\includegraphics{#2.pdf}}}{#3}{#4\vspace{#1}}}
\newcommand{\eql}[1]{\begin{equation}#1\end{equation}}

\newcommand{\eqi}[1]{$#1$}

\DeclareRobustCommand*{\unit}[1]{~\ensuremath{\mathrm{#1}}}

\RequirePackage{color}
\definecolor{darkgreen}{rgb}{0,0.45,0}

\input{Journal.inc}

\begin{document}

\setlength{\topmargin}{-20pt}
\AANum{A72}
\yearCop{2015}
\doi{\doiurl{10.1051/0004-6361/201525839}}
\idline{A\&A 580, A72 (2015)}
\hypersetup{
	pdftitle = {Coronal energy input and dissipation in a solar active region 3D~MHD model},
	pdfauthor = {Ph.-A.~Bourdin, S.~Bingert, and H.~Peter},
	pdfkeywords = {Sun: corona -- magnetohydrodynamics (MHD) -- methods: numerical -- Sun: UV radiation},
	pdfsubject = {Astronomy \& Astrophysics}
}

%
\title{Coronal energy input and dissipation in a solar active region 3D~MHD model}
\titlerunning{Coronal energy input and dissipation in an AR model}

\author{Ph.-A.~Bourdin\inst{1,2}, S.~Bingert\inst{3}, and H.~Peter\inst{2}}
\authorrunning{Ph.-A.~Bourdin et al.}

\institute{%
Space Research Institute, Austrian Academy of Sciences, Schmiedlstr. 6, 8042 Graz, Austria
\\
\email{Philippe.Bourdin@oeaw.ac.at}
\and
Max-Planck-Institut f{\"u}r Sonnensystemforschung, Justus-von-Liebig-Weg 3, 37077 G{\"o}ttingen, Germany
\and
Gesellschaft f{\"u}r wissenschaftliche Datenverarbeitung, Am Fa{\ss}berg 11, 37077 G{\"o}ttingen, Germany
\vspace{6pt}
}

\date{Received 7 February 2015 / Accepted 18 June 2015}

\abstract%
{%
We have conducted a 3D MHD simulation of the solar corona above an active region (AR) in full scale and high resolution, which shows coronal loops, and plasma flows within them, similar to observations.
}
{%
We want to find the connection between the photospheric energy input by field-line braiding with the coronal energy conversion by Ohmic dissipation of induced currents.
}
{%
To this end we compare the coronal energy input and dissipation within our simulation domain above different fields of view, e.g. for a small loops system in the AR core.
We also choose an ensemble of field lines to compare, e.g., the magnetic energy input to the heating per particle along these field lines.
}
{%
We find an enhanced Ohmic dissipation of currents in the corona above areas that also have enhanced upwards-directed Poynting flux.
These regions coincide with the regions where hot coronal loops within the AR core are observed.
The coronal density plays a role in estimating the coronal temperature due to the generated heat input.
A minimum flux density of about 200\unit{Gauss} is needed in the photosphere to heat a field line to coronal temperatures of about 1\unit{MK}.
}
{%
This suggests that the field-line braiding mechanism provides the coronal energy input and that the Ohmic dissipation of induced currents dominates the coronal heating mechanism.
}
\keywords{ Sun: corona -- magnetohydrodynamics (MHD) -- methods: numerical -- Sun: UV radiation \vspace{6pt}}

\maketitle

\section{Introduction\label{S:energy.intro}}

The coronal heating mechanism is poorly understood, although various theories exist -- many of which are listed in the review of \cite{Klimchuk:2006}.
A coronal heating mechanism including the actual dissipation is field-line braiding due to horizontal shuffling motions in the photosphere that lead to induced currents in the corona where these currents heat the plasma though Ohmic dissipation \citep{Parker:1972}.
Models of solar active regions indicate that this mechanism can produce a loop-dominated corona \citep{Gudiksen+Nordlund:2002,Gudiksen+Nordlund:2005a,Gudiksen+Nordlund:2005b}.

Statistical analyses of synthesized spectra from these models match statistical properties of observed spectra \citep{Peter+al:2004,Peter+al:2006}.
The Doppler shifts in these models were able to reproduce the observed persistent redshifts in the transition region, but failed to show the observed blueshifts at higher temperatures above 500\,000\unit{K} \citep{Peter:1999full,Peter+Judge:1999}.
This is also true for more recent active region models \citep{Zacharias+al:2011a,Zacharias+al:2011b}.
Though quiet Sun (QS) models do show blueshifts at high temperatures, the magnitudes of their red- and blueshifts are too small \citep{Hansteen+al:2010}.
We present here a model of an active region in a full-scale computational domain that implements the field-line braiding process and that is able to reproduce the redshifts in the transition region and the blueshifts in the corona.

We can follow the magnetic field through the model corona and extract any quantity along that field line, which is otherwise inaccessible by coronal observations.
We find that a minimum magnetic flux density of 200\unit{Gauss} is necessary at the footpoints of these loops to produce a sufficient upwards directed Poynting flux to heat the coronal part of these loops by Ohmic dissipation.

\section{Coronal model and properties of energy input\label{S:model}}

In our study \citep{Bourdin+al:2013_overview}, we have conducted a 3D magnetohydrodynamical (MHD) numerical simulation of the solar corona above an active region (AR).
The model includes gravity, radiative losses following \cite{Cook+al:1989}, field-aligned heat conduction \citep{Spitzer:1962}, and a uniform magnetic diffusivity \eqi{\eta = 10^{10}} \unit{m^2/s}.
The plasma pressure and the magnetic field in the corona are described self-consistently.
In order to do this, it is important to set up a realistic energy balance, including necessary coronal energy sinks.
The Spitzer-type heat conduction transports energy downwards to the chromosphere and also powers evaporation of colder plasma there.

\graphwidth[18.6cm]{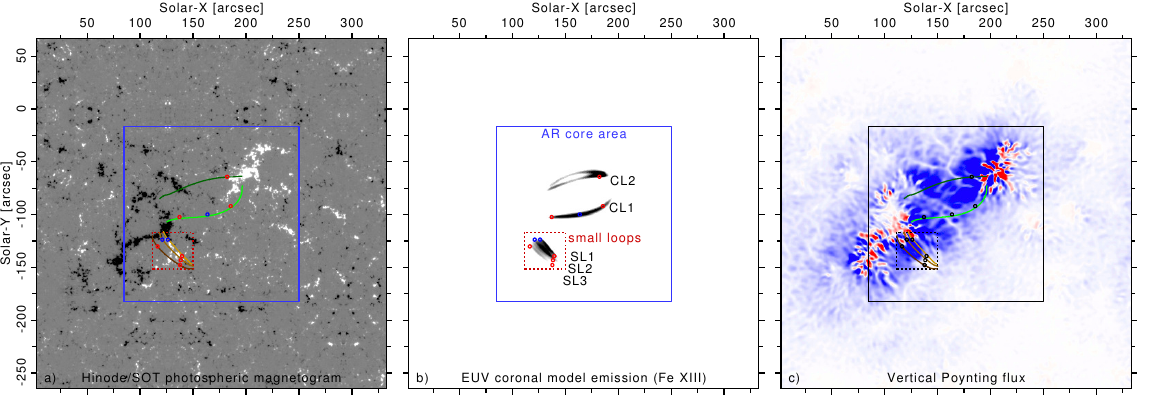}{model.overview}{
Active region observed by the Hinode satellite on 14\,Nov\,2007.
\panel{Left} line-of-sight magnetogram (saturation level: \eqi{\pm\,300\unit{G}}) of the AR including magnetic field lines along the most prominent coronal loops \citep[CL\,1--2 and SL\,1--3, see][]{Bourdin+al:2013_overview}.
\panel{Middle} \ion{Fe}{13} EUV emission synthesized from the model data.
The footpoints of the loops are marked with circles, where blue indicates plasma upflows and red downflows.
\panel{Right} vertical component of the Poynting flux at 3\unit{Mm} height.
Upwards oriented vertical flux is colored in blue and downwards flux in red, both in linear scale with a saturation level of \eqi{\pm\,50\,000\unit{W/m^2}}.
The solid square indicates the hot AR core area, while the dotted rectangle outlines the FOV of only the short loops SL\,1--3, see also \sect{energy.source}.}

We drive the model from below by photospheric observations, using Hinode/SOT-NFI \citep{Kosugi+al:2007,Tsuneta+al:2008} line-of-sight (LOS) magnetograms and horizontal velocities.
Large-scale motions of magnetic patches and granulation-like small-scale velocities introduce magnetic stress at the magnetic field footpoints.
These perturbations propagate along the field from the photosphere into the corona and carry magnetic energy as Poynting flux.
Ohmic dissipation of induced electric currents turned out to be the main contribution to the model corona heating mechanism.

With the atomic database {\sc Chianti} \citep{Dere+al:1997,Young+al:2003} we deduce the synthetic emission following \cite{Peter+al:2004,Peter+al:2006} and compare it to real observations of the same active region, like the EUV and X-ray emission spectra from highly ionized atoms and their corresponding Doppler line-shifts, taken by Hinode/EIS \citep{Culhane+al:2007}.
We found a substantial match of coronal structures and their flow dynamics between the simulated and observed loops system.

\subsection{Boundary conditions\label{S:boundary}}

Our model is periodic in the horizontal directions.
The lower and upper boundaries of the computational domain are closed to any plasma inflow or outflow, and the temperature gradient is zero at the top and bottom boundaries to avoid any artificial energy flow into or out of the domain.
The density is set according to the temperature, so that the boundary of the physical domain is in hydrostatic equilibrium.
We extrapolate the magnetic field at the upper boundary with a potential field into the three ghost layers.

For the magnetic field at the lower boundary, we use the potential-field extrapolation into the interior of the Sun, which is in principle identical to a potential-field extrapolation to the exterior, except that we do not decrease the contrast exponentially depending on the distance from the physical boundary layer; instead, we increase the contrast of the vertical magnetic field with depth.
The pressure scale height below the photosphere is about 300\unit{km}, but for numerical reasons this scale height is stretched in our boundary condition to about 1000\unit{km}.
On the other hand, for the relatively short distance of about 300\unit{km} that we need to extrapolate into the solar interior, this choice seems to be relatively safe because the typical length-scales of advectional structures in the upper convection zone are usually larger than 1000\unit{km} in the vertical direction.
Furthermore, a small error in the horizontal field component at the photospheric level would have almost no influence on our model corona.

\subsection{Initial condition and switching on\label{S:initial}}
We start with a stratified atmosphere that we combined from temperature and density profiles of the solar interior \citep{Stix:1989}, the chromosphere \citep{Fontenla+al:1993}, and the corona \citep{November+al:1996}.
We relax this combined profile into equilibrium using a 1D hydrodynamic model of a vertical atmospheric column together with a velocity damping and the same diffusion parameters as we use for the 3D MHD model.

Granulation has a typical life time of about 5\unit{min} and the Alfv{\'e}n crossing time of the magnetic energy input from the photospheric driving to reach the corona is about 30\unit{min}.
From minute 10 to 35, we smoothly switch on the radiative losses and the Spitzer-type heat conduction, to compensate for a lack of coronal heating during this initial phase.
All physical processes of the model are fully active after minute 35.
To damp switch-on effects in the 3D MHD model, we use a velocity damping that fades out smoothly until minute 40.

After the Alfv{\'e}n crossing time plus about three granular life times, we see a strong rise in the coronal Ohmic heating around minute 48 in the simulation.
At that time, the maximum temperature from the initial atmosphere has dropped from 1.2\unit{MK} to about 0.4\unit{MK} and then rises again to about 1.4\unit{MK} until \mbox{minute 60}, where it reaches a kind of plateau.
The small loops system (SL\,1--3, see \sect{overview.box}) is developed at minute 63 and was then stable in the model for about 7 solar minutes.

\section{Energy input viewed at large\label{S:energy.input.large}}

\subsection{Overview of whole box and subvolumes\label{S:overview.box}}
The AR we investigate in this work has a size of about \eqi{200\times100\unit{Mm^2}}.
Observations exist from the photosphere, in particular LOS magnetograms and calibrated vector magnetograms of a smaller field of view (FOV); coronal observations were also made co-spatial and co-temporal in the range of X-rays and extreme ultra violet (EUV), see \firstfig{model.overview}.
For the simulation we use the full FOV shown in \fig{model.overview}, including some surrounding QS, so that effects of the horizontal periodicity within our model are minimized.
Mainly, the FOV can be divided into three different regions: the full FOV including the QS, the AR core area, and the small loops system within the AR core, see \fig{model.overview}.
While restricting our analyses to the small loops system (SL\,1--3) for example, we can study a region with enhanced coronal energy dissipation and strong EUV emission.

\graphreduced{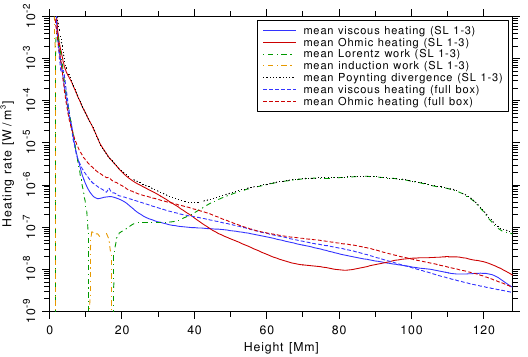}{heating_profile}{
Horizontal averages of the volumetric Ohmic (red) and viscous heating (blue) plotted as a height profile, where the line styles (solid and dashed) depict the different FOVs given in brackets in the legend.
The dash-dotted lines basically describe the same physical quantity \eqi{W_{Lorentz} = \vec{u} \cdot \left(\vec{\nabla} \times \vec{B} \times \vec{B}\right) / \mu}, while the Lorentz work (green) and the induction work (orange) are of opposite sign.
Due to energy conservation, the sum of all these quantities must be equal to the Poynting flux divergence, e.g. for the SL\,1--3 FOV (black dotted).}

\subsection{Contribution to the internal energy\label{S:internal.energy}}

To investigate the spatial distribution of the energy input into the model corona, we look at the individual heating terms in a horizontal average at a given height.
The driving in the bottom boundary layer provides the energy input by Poynting flux into the model.

In \firstfig{heating_profile} we show which energy dissipation mechanism is active at which height for different fields of view (FOV).
While below the corona all profiles lie relatively close together, looking at the full FOV data (dashed lines) we see that in the corona the Ohmic heating dominates over the viscous heating at almost all heights.
In general, the coronal part of the profiles drops roughly exponentially with height.
This applies for the box average as well as for individual loops (SL\,1--3).
The three loops SL\,1--3 show a very similar variation of the heat input along the field lines (and thus with height), which is why we show here only one curve averaging the three loops.
The result of the energy input dropping with height for individual field lines is consistent with \cite{vanWettum+al:2013} who found that the volumetric heating drops exponentially with height also for individual field lines.

The scale height of the Ohmic and viscous heating (averaged over the full box FOV) are very similar and uniform in the whole corona.
When we restrict the FOV to the hot AR core loops SL\,1--3 (solid lines) the Ohmic heating profile gets more structured.
Between the heights of 5 and 30\unit{Mm}, where the hot loops are located, we see a significant increase of the Ohmic heating compared to the full FOV.

In addition, we see a shortage of the Ohmic heating above the hot AR core loops in heights between 40\unit{Mm} and 90\unit{Mm} in relation to the full FOV average (cf. \fig{heating_profile}).
This can be understood by assuming that the energy transport is directed mainly upwards and that an increased dissipation in a lower layer would decrease the possible energy input to an upper layer.
Of course there is more to consider; for example, most of the field lines that close well above the hot AR core subvolume are rooted outside that area, so that energy can be transported there while following the magnetic field.

Both the full and reduced FOV profiles in \fig{heating_profile} approximately join up again at heights above 90\unit{Mm}, which indicates that either some energy is transported to above the hot AR core loops subvolume from outside.
Alternatively, this upper part of the corona could simply not yet have become fully independent of the initial condition because the Alfv\'en travel time for magnetic perturbations from the bottom layer into the upper corona is significantly longer than into the hot AR core loops system.

The profile of the divergence of the Poynting flux mainly follows the sum of the Ohmic heating and the Lorentz work (see \fig{heating_profile}).
Between 10 and 20\unit{Mm} we find a region that is in average dominated by induction work.
For the full FOV average, this induction region would expand farther from 10 to about 55\unit{Mm}.

The viscous heating also shows deviations between the two FOV averages, but the difference is much smaller than the deviations seen in the Ohmic heating (red and blue lines in \fig{heating_profile}).
Also for the small FOV, the viscous heating is weaker between heights of about 10 to 100\unit{Mm} than for the full FOV.
Above 6\unit{Mm} height the viscous heating in the small FOV (containing SL\,1--3) is always at least a factor of 5 weaker than the Ohmic heating or the Lorentz work.
Therefore, the viscous heating only provides a minor contribution to dissipating the Poynting flux in the corona.
Altogether, this shows that the heat input to the hot AR core loops in our model is largely dominated by the Ohmic heating.

\subsection{Contribution to the temperature\label{S:temperature.profile}}

\graphreduced{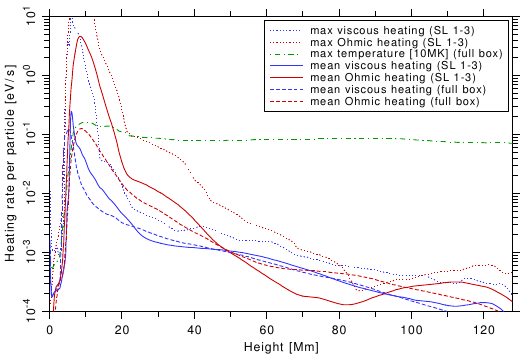}{heating_particle_profile}{
Similar to \fig{heating_profile}, but for the heating per particle.
Horizontal averages of Ohmic (red) and viscous heating per particle (blue) versus height.
The dotted curves indicate the maximum values at each height, while the solid lines show the corresponding mean value.
Overplotted is the height profile of the maximum temperature (green dash-dotted).
The different FOVs are symbolized as line styles and are given in brackets in the legend.}
The question of whether the viscous heating has a significant impact on the overall temperature stratification remains.
To address this question, one must also consider the plasma density at the places where the heating occurs in order to understand how much this volumetric heating can increase the local temperature.
We show this in \firstfig{heating_particle_profile} by looking at the Ohmic and viscous heating rates per particle.
We find that the mean Ohmic heating per particle (red) dominates over the viscous heating (blue) already for the full FOV (dashed lines) from a height of 8\unit{Mm} upwards.

When restricting the FOV to the hot AR core loops \mbox{SL\,1--3} (solid lines in \fig{heating_particle_profile}), we see a significant increase of the Ohmic heating per particle in the heights between 5 and 40\unit{Mm}, where these loops are located.
We also see a slight increase of the viscous heating per particle at these heights, but in the upper corona the viscous heating per particle remains roughly unchanged.
Above the smaller hot AR core FOV, the Ohmic heating clearly dominates over the viscous heating by a factor of 10 and more up to a height of roughly 30\unit{Mm}.
Furthermore, in this region the mean of the Ohmic heating per particle (red solid line) also reaches over any local maximum of the viscous heating (blue dotted line).
In the upper corona above 50\unit{Mm}, we see the same tendency as in \fig{heating_profile}, namely that the Ohmic heating per particle significantly drops below the viscous heating for the restricted FOVs, while this is not the case for the full FOV (dashed lines).

The viscous heating can gain dominance over the Ohmic heating above hot closed-field regions that are strongly heated by Ohmic dissipation.
We see that the viscous heating is slightly increased above the crossing point of Ohmic and viscous heating compared to the full box FOV, while it is slightly decreased below this crossing point.
In contrast to the restricted FOV, the global mean viscous heating profile has a rather constant scale height above 20\unit{Mm}.
This behavior of the viscous heating could be explained by downflows from the upper corona that are disturbed by the magnetically closed AR core.
Above 80\unit{Mm} the height profiles of the Ohmic heating per particle dominate again over those of the viscous heating with a factor larger than 2.

Nonetheless, the maximum temperature stratification remains mainly constant in the corona above 20\unit{Mm}.
The rapid rise of this temperature profile coincides exactly with the strong rise in Ohmic and viscous heating in the lower 3-7\unit{Mm}.
The strong gradient of the temperature ends exactly where the strong increase of the viscous heating ends and does not follow the additional rise in the Ohmic heating towards 10\unit{Mm}.

This overview indicates that the main contribution to the high coronal plasma temperatures is by the Ohmic heating, while the viscous heating seems to contribute mainly in higher density areas where the local input to the internal energy would not result in such a high temperature rise.
Still, the location and shape of the transition region temperature rise seems to be determined by the Ohmic and viscous heating equally, where both of them deliver similar amounts of internal energy.

We find a peak in viscous heating per particle at the base of the transition region (around 6\unit{Mm}), where coronal downflows (of approximately 200 km/s) are practically stopped at the top of the chromosphere by momentum conservation.
The strong velocity gradients on the grid scale are numerically stable over the whole simulation run.
These downflows are caused partly by cooling material converting potential into kinetic energy.
In addition, some material that was previously accelerated by the Ohmic heat input leads to an increase in the plasma pressure and hence pushes material via adiabatic expansion.
Therefore, some of the later viscous heating is ultimately caused by the initial Ohmic dissipation.

The location of the strongest Ohmic heating per particle along individual loops can be inferred from \fig{heating_particle_profile}, where the red~solid line represents the average over the volume containing the three smaller loops SL\,1--3 that show the strongest heating rates at a height of 8\unit{Mm} and have apex heights of 11--13\unit{Mm}.
Furthermore, the hot and larger loop CL\,1 has its strongest heating per particle between 6 and 7\unit{Mm} and reaches an apex height of 17\unit{Mm}, while the much cooler and less EUV-emissive loop CL\,2 is 21\unit{Mm} high but its heating also peaks around 6 to 7\unit{Mm} in the lower corona.
The heating per particle is obviously strongest at some intermediate height, which we would like to refer to as the ``legs'' of the loops.

\subsection{Energy supplied into the model domain\label{S:energy.source}}
Perturbations in the magnetic field can travel into the corona with the Alfv\'en velocity.
Depending on the speed of the driving motions, these perturbations will follow the magnetic field as transversal or longitudinal waves that could be dissipated in the solar atmosphere, which is referred to as ``AC'' heating.
If the driver only gradually changes a field line, currents can be induced that compensate these slow perturbations, which can be seen as a diffusional process that belongs to ``DC'' heating \citep[e.g.,][]{Klimchuk:2006}.

The field-line braiding introduced by our lower boundary creates the Poynting flux that powers the coronal energy input.
In the photosphere, the applied horizontal motions shuffle the footpoints of magnetic field lines with velocities of about 1\unit{km/s}.
This bends the magnetic field that starts to deviate from the initially force-free state and these perturbations can then propagate with roughly the Alfv\'en speed along the field lines.
As a result, a net upwards directed Poynting flux carries magnetic energy from the photosphere into the corona, where it can then be dissipated.
In the right panel of \firstfig{model.overview} we display the vertical Poynting flux at the base at the corona, which is located at a height of about 3\unit{Mm}.

The vertical component of the Poynting flux is highly structured at the photospheric level, as are the magnetic field and the horizontal velocities.
A change of the photospheric advection direction of the magnetic field can already cause a sign reversal in the vertical Poynting flux.
This is why we see strong fluctuations (blue and red stripes in \fig{model.overview}\,c) above the main polarities in the AR core even at the base of the corona.

With increasing height, the magnetic field is smoothed out by the magnetic expansion and the contrast in the Poynting flux also smoothes quickly.
In order to become independent of the large fluctuations in the photosphere, and also to have a reliable and smooth estimate of the magnetic field lines, we choose the height where the temperature reaches a value of 75\,000\unit{K} as point of reference for the footpoints of a field line.
This temperature lies above the chromosphere but still within the transition region that is typically located at heights between 3\unit{Mm} and 5\unit{Mm} in our model.

An observed coronal loop is typically a roughly semi-circular structure with a field inclination of 0 to 30 degrees in the lower 3\unit{Mm} of the atmosphere, i.e., below the transition region.
Because the magnetic field can vary strongly there, a good estimate of the loop length \eqi{L} would disregard a field line that expands to remote regions by following directional changes in the chromospheric magnetic field.
For this analysis, the loop length \eqi{L} is defined as the distance between both loop footpoints plus the distance to the photosphere, which we compute by assuming an average inclination of 20 degrees for the magnetic field below the transition region.
Only then can we compare our model data to scaling laws that are usually based on observations and only use the visible coronal part of a loop to determine its length.
If we use another average inclination between 0 and 30 degrees below the transition region, the results of this analysis deviate only marginally because the loop length is dominated by the coronal part.

\section{Statistical analyses of coronal field lines\label{S:statistical.analyses}}
For a statistical analysis of coronal field lines we first need to select a sample of field lines that represent the AR corona.
From our model we extract not only the field lines along hot coronal loops, but also field lines rooted all over the AR region core.
The seed points for the field-line tracing are distributed along a vertical cut through the domain, located roughly in the middle between the two main polarities of the AR.
Selected field lines are a) longer than 18\unit{Mm} in total; b) shorter than 150\unit{Mm}; and c) reach at least to the given minimum temperature of 75\,000\unit{K} somewhere below 18\unit{Mm} height.
If a field line does not fulfill one of these conditions it either closes below the corona, it is located in a cool quiet region and would not produce coronal emission in the spectral lines we are interested in, or it spans high up into the corona where the magnetic perturbations from the bottom layer might not yet have sufficiently propagated.
We also exclude open field lines from the analysis.
In addition to this rather ``global'' ensemble, for each hot AR core loop we added one field line that crosses the maximum intensity of that loop, as well as about 200 field lines neighboring these loops (see \fig{model.overview}).
The 67\,000 field lines that we use for this analysis have apex heights of about 6 to 80\unit{Mm}, while the most prominent loops seen in synthesized EUV emission reach heights between 10 and 22\unit{Mm}.

\subsection{Coronal energy conversion\label{S:energy.conversion}}
In \sect{energy.input.large} we analyzed the coronal energy conversion process only with spatial averages over different subvolumes.
In this section we now turn to the coronal energy conversion process, in particular the Ohmic heating and its relation to the photospheric driving and to coronal temperatures, by looking at ensembles of coronal field lines.

\subsubsection{Poynting flux as energy source\label{S:poynting.flux}}

We investigate the influence of the Poynting flux in the transition region on the coronal energy input.
The Poynting flux into a field line is computed as the sum of the directed flux at both field-line footpoints, while an upward flux is positive and a downward flux is negative.
This sum gives a measure for the net vertical Poynting flux that actually remains in the corona, if that value is positive (58\,000 field lines).
For the one snapshot we use for this study, about one seventh of the selected field lines show a negative net flux (9000 field lines), which stands for a magnetic energy outflow from the corona downwards.
The sign of the vertical Poynting flux will fluctuate over time, depending on the evolution of the photospheric driving motions.
Nonetheless, also for a single snapshot, the horizontal average of the divergence of the Poynting flux is positive at all heights within the hot AR core loops FOV, see \fig{heating_profile}.

In \firstfig{P_z_HR_ohm} we relate the vertical net Poynting flux into a field line to the volumetric Ohmic heating integrated between both field-line footpoints, not taking into account the part below the transition region.
The main bulk of the population is of positive net Poynting flux (dark blue).
Negative net Poynting flux also occurs (red), but the strongest Poynting fluxes are all positive, as can be seen at the right end of the distribution in \fig{P_z_HR_ohm}.
The large scatter is due to the evolution of the photospheric driving where we look only at one snapshot in time.
A negative net Poynting flux (red points in \fig{P_z_HR_ohm}) can be explained, for example, by coronal induction or Lorentz work that increases the magnetic energy locally, which can then be transported downwards.
Another likely process is that the negative flux (produced at the lower boundary) cancels out with positive flux from neighboring field lines due to the magnetic expansion in the corona.
In total, the mean Poynting flux for the full FOV of our model is nonetheless positive at every height.

All hot AR core loops (marked by symbols) lie within the positive net flux population on the upper right except for SL\,3, which has a negative Poynting flux, see \sect{Ohmic.mean}.

We fit a scaling law with an exponent of \eqi{\alpha = 0.47\,\pm\,0.06} to the positive data points (see \fig{P_z_HR_ohm}).
The error is computed from the standard deviation of the LAD fit and by estimating the width of the data distribution to four orders of magnitude in the horizontal direction.
While the scatter of the individual data points is quite large, the trend found by the fit is still clear.
This is supported by grouping the data points into bins of the vertical Poynting flux (x-axis in \figs{P_z_HR_ohm}{P_z_HR_ohm_part_mean}) and calculating the mean of the Ohmic heating in these bins.
These data points are shown by green stars in \figs{P_z_HR_ohm}{P_z_HR_ohm_part_mean}.
This shows a small scatter and a clear trend consistent with the power-law fit.
The positive slope of the fit is consistent with a relation between the volumetric Ohmic heating integrated along a coronal field line and the vertical Poynting flux \eqi{{\cal{P}}_z} at the field-line footpoints of the form
\eql{F_{Ohm} \sim {{\cal{P}}_z}^\alpha \label{E:F_Ohm_P_z} ,}
where \eqi{\vec{\cal{P}} = \eta \left(\vec{\nabla} \times \vec{B}\right) \times \vec{B} + \left(\vec{u} \times \vec{B}\right) \times \vec{B} / \mu} with the velocity \eqi{\vec{u}}, the magnetic field vector \eqi{\vec{B}}, the magnetic diffusivity \eqi{\eta}, and the permeability \eqi{\mu}.
Above a height of 2\unit{Mm}, where all the footpoints of our field lines are located, the Poynting flux is largely dominated by the second term that reflects transversal perturbations in the magnetic field.
As the main distribution of data points is clearly located below the equality line, we conclude that not all of the Poynting flux in our model is usually converted to Ohmic heat.
The hot AR core loops SL\,1--3 and CL\,2 convert a larger part of their incoming Poynting flux into Ohmic heat than the average field lines do.

\graph{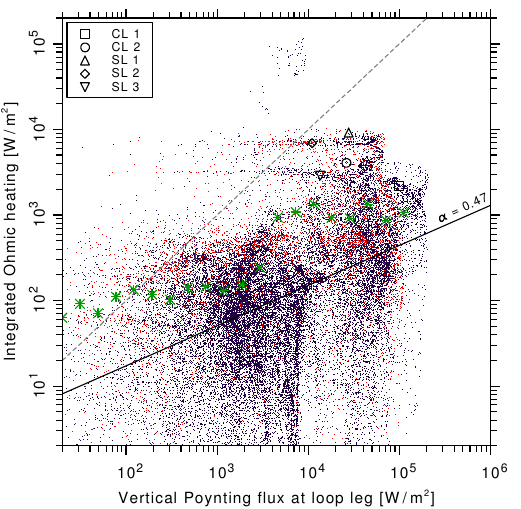}{P_z_HR_ohm}{
Vertical net Poynting flux into a field line versus the volumetric Ohmic heating integrated along that field line.
Red data points show the absolute value of the Poynting flux wherever the Poynting flux is negative, dark blue dots show the data points where the Poynting flux is positive.
The gray dashed line indicates the equality, where all of the incoming Poynting flux would be converted to Ohmic heat.
We apply a least-absolute-deviation (LAD) fit to the positive net flux data points (black line).
The green stars indicate the mean Ohmic heating within the corresponding bins in Poynting flux.
See \sect{poynting.flux}.}

\graph{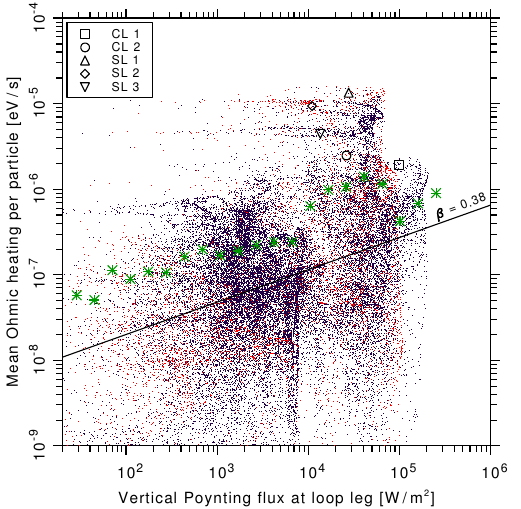}{P_z_HR_ohm_part_mean}{
Same as \fig{P_z_HR_ohm}, except that we show here the mean Ohmic heating per particle along a field line.}

Generally, the Ohmic heating integrated along field lines is smaller than the Poynting flux, although the energy is conserved because not all Poynting flux needs to be converted into Ohmic heat (or dissipated at all).
Some of the remaining upwards magnetic energy transport that is not dissipated can leave the simulation box though the upper boundary or it can be converted into magnetic field stress that can let the coronal magnetic field evolve away from a potential field configuration, or it can lead to a rise of magnetic structures into the upper corona.
For instance, we found that the large AR core loops CL\,1 and 2 are still rising with about 2\unit{km/s} at the loop top \citep{Bourdin+al:2013_overview}.
Accordingly, we find a net upwards mass transport in the middle of the AR core between the two main polarities.

We now perform the same analysis with the mean Ohmic heating per particle instead of the volumetric heating, see \firstfig{P_z_HR_ohm_part_mean}.
Again, we overplot a LAD fit to a scaling law that gives an exponent of \eqi{\beta = 0.38\,\pm\,0.07} (black dotted line), which indicates a weaker correlation.
We also find a 14\unit{\%} higher variance in the logarithmic data, and hence we see more scatter in \fig{P_z_HR_ohm_part_mean} than in \fig{P_z_HR_ohm} because the heating per particle also depends on the particle density \eqi{n}, which plays a role when relating the magnetic energy input not only to the dissipated magnetic energy, but also to the actual heating of the coronal plasma.

For the strongest heated loops SL\,1+2 we find a heating per particle of about \eqi{10^{-5}\unit{eV/s}}, while the average for field lines with the same magnetic energy input is about 10 times smaller (see the green stars around \eqi{10^{-6}\unit{eV/s}} below the black symbols in \fig{P_z_HR_ohm_part_mean}).

In both \figs{P_z_HR_ohm}{P_z_HR_ohm_part_mean}, the hot loops (symbols) clearly have a stronger heating than the scaling law fitted through all data points would predict (black line) and also a higher temperature than the average of field lines with a similar Poynting influx (green stars).
All in all, this is good because positive exponents \eqi{\alpha} and \eqi{\beta} support a possible correlation between the net Poynting flux into a field line and its temperature, via the Ohmic heating in the corona, which we would expect.

We get a similar result when we use the Ohmic heating per particle instead of the volumetric heating for this analysis.

\subsubsection{Ohmic heating integrated along field lines\label{S:Ohmic.integrated}}

For the same ensemble of field lines as in \sect{poynting.flux}, we investigate in \firstfig{HR_ohm_total_Temp_max} the Ohmic heating per particle.
We find here that the population of hot field lines mainly follows a scaling law of the form
\eql{T_{max} \sim {F_{Ohm}}^{2/7} \label{E:T_F_scaling} ,}
where the maximum temperature \eqi{T_{max}} relates to the Ohmic heating integrated along that field line \eqi{F_{Ohm}} that we calculate from the volumetric Ohmic heating \eqi{H_{Ohm}} as:
\eql{F_{Ohm} = \int {H_{Ohm}(s) \cdot ds} \label{E:F_Ohm} .}
The exponent of \eqi{2/7} is predicted by the RTV scaling law \citep[named after the authors of][]{Rosner+al:1978}.

While the distribution of data points in \fig{HR_ohm_total_Temp_max} is relatively narrow for the high heating rates and high temperatures, the distribution becomes much broader for field lines with a maximum temperature below 1\unit{MK}.
These cooler field lines have insufficient heating to become hot and are not quasi-static.
Therefore, these will not follow the RTV scaling law and one expects scatter.

However, the hot end of the population of field lines with more than 100\unit{W/m^2} of integrated Ohmic heating follows the slope of \eqi{2/7} and is comparable with a scaling law, see upper left bound in the right half of \fig{HR_ohm_total_Temp_max}, which represents the hot coronal loops that the RTV scaling was originally designed for.

Still, the large scatter in \fig{HR_ohm_total_Temp_max} shows that the scaling laws of \cite{Rosner+al:1978} can only be considered an average variation.
Any realistic model considering the temporal variability and the large range of heating rates will produce loops that do not follow the RTV laws.
In this sense the RTV scaling laws should be considered only on a statistical basis.

\graph{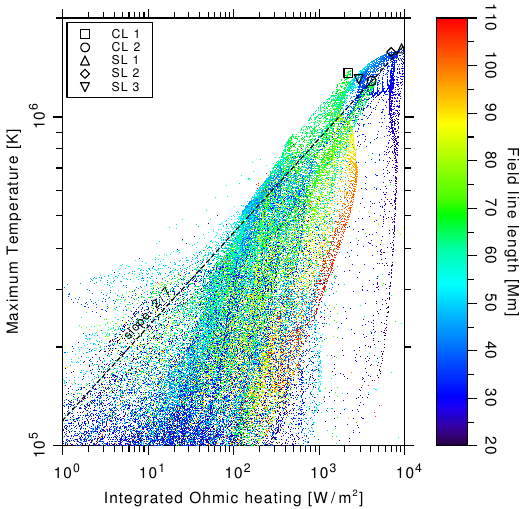}{HR_ohm_total_Temp_max}{
Volumetric Ohmic heating integrated along a field line (\eqi{F_{Ohm}}) versus its maximum Temperature (\eqi{T_{max}}).
The hottest loops (marked with symbols) all have a high Ohmic heating.
The three loops with the strongest EUV emission (SL\,1+2 and CL\,1) lie exactly at the local maxima of their surrounding population.
The dashed black line indicates the exponent of \eqi{2/7} with the parameters of the loop SL\,1 as anchor point, see \sect{Ohmic.integrated}.}

\graph{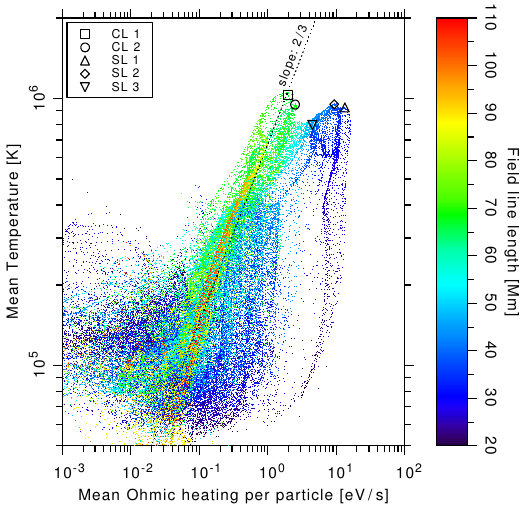}{HR_ohm_part_mean_Temp_mean}{
Mean Ohmic heating per particle along a field line (\eqi{\langle H_{Ohm} / n \rangle}) versus its mean Temperature (\eqi{\langle T \rangle}).
The mean is computed here as the heating integrated along a field line divided by the integration path length.
The field lines with a heating rate above 10\eqi{^{-1}\unit{eV/s}} are roughly ordered according to their length.
The black dotted line indicates a slope of \eqi{2/3}, see \sect{Ohmic.mean}.}

\subsubsection{Mean Ohmic heating\label{S:Ohmic.mean}}

Now we investigate the relation between the average heating per particle and temperature along a field line.
For that, we need to replace the volumetric Ohmic heating \eqi{H_{Ohm}} in \eqn{F_Ohm} with the Ohmic heating per particle \eqi{H_{Ohm}/n}.
The RTV scaling law together with the ideal gas law \eqi{p \sim n k_B T} predicts a slope of \eqi{2/3}:
\eql{\langle T\rangle \sim {\langle H_{Ohm} / n\rangle}^{2/3} \label{E:T_mean_H_part_mean_scaling}}

The relation in \firstfig{HR_ohm_part_mean_Temp_mean} between the mean Ohmic heating per particle and the mean temperature along a field line is clearly visible for field-line lengths between 60 and 110\unit{Mm} (turquoise to red colors), which follow nicely the predicted slope of \eqi{2/3} (black dotted line).
For field-line lengths below 55\unit{Mm}, as well as for mean temperatures below 0.2\unit{MK}, we find large scatter.

Nonetheless, here we see a clearer match to the predicted slope, even for groups of different field-line lengths above 60\unit{Mm} (see red, yellow, green and turquoise populations in \fig{HR_ohm_part_mean_Temp_mean}), than in \fig{HR_ohm_total_Temp_max}.
This shows again, as discussed in \fig{P_z_HR_ohm_part_mean}, that the density is a relevant ingredient in order to describe not only the coronal heating, but also the resulting plasma temperature along a field line.

The short and hot loops SL\,1+2 show the strongest mean heating, while the longer loops CL\,1+2 have a 10 times lower average heating, even though their lengths (35 and 70\unit{Mm}) differ only by a factor of 2.
Still, they have roughly the same average temperature, because the very efficient heat conduction and the radiative losses act like a thermostat.
We see the same results when we look at the volumetric heating (\fig{HR_ohm_total_Temp_max}) instead of the heating per particle.

The loop SL\,3 is relatively cool and does not show much emission in EUV and X-rays and is therefore not a typical hot coronal AR loop \citep[cf.][]{Bourdin+al:2013_overview}.
The lower mean Ohmic heating can be explained because SL\,3 is the only loop with a negative net Poynting flux, see \sect{poynting.flux}.
CL\,2 is not a developed hot loop and lies outside the main coronal emission structures.

CL\,1 lies farther away from the equality line (gray line in \fig{P_z_HR_ohm}), which indicates that this loop uses a smaller fraction of the incoming Poynting flux for Ohmic heating than the other hot AR core loops.
Owing to the particularly high incoming \mbox{Poynting} flux, CL\,1 still can reach Ohmic heating rates as high as the other loops.

\subsubsection{Relation to magnetic field strength\label{S:magnetic.field.relation}}
Also of high interest is the connection between the field-line length and the total Ohmic heat input (which sets a field line's tendency towards high or low coronal temperature) in relation to the magnetic flux density at the footpoints of the field line.
As mentioned in \sect{boundary}, there are strong fluctuations of the vertical magnetic flux density in the photospheric layer.
It is important that this analysis is independent of these strong fluctuations that would anyway not reach into the corona.
Therefore we choose 300\unit{km} as the reference height for the magnetic flux into a field line, which is roughly one photospheric pressure scale-height above the photosphere, but still below the chromosphere in our model.

\graph{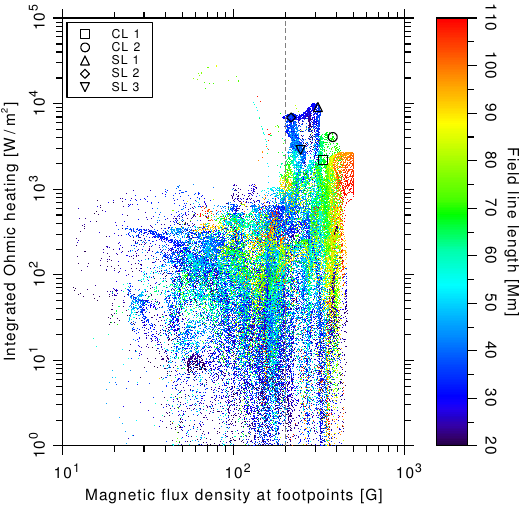}{B_abs_HR_ohm_total}{
Magnetic flux density at the footpoints located at z=300\unit{km} height versus volumetric Ohmic heating integrated along the field line.
The hot loops SL\,1+2 lie within the high heating rates population.
The vertical gray dashed line indicates the magnetic flux density threshold of 200\unit{G}, see \sect{magnetic.field.relation}.}

In \firstfig{B_abs_HR_ohm_total} we display the mean of the vertical magnetic flux density values at both field-line footpoints versus the volumetric Ohmic heating integrated along each field line.
The field-line length (color coded) is taken between the reference heights where the temperature reaches 75\,000\unit{K}, plus the extension towards the photosphere.

Below a critical magnetic flux density of roughly 200\unit{G} (gray dashed line in \fig{B_abs_HR_ohm_total}), we find a wide distribution of flux densities with no general correlation to the integrated Ohmic heating or to the field-line length.
Nevertheless, almost all strongly heated field lines are located above this critical flux density.
We do not see a strong correlation between the footpoint flux density and the coronal heating rate, but the field-line lengths clearly separate and range from the long field lines (100\unit{Mm}, red at the lower right) with relatively low Ohmic heating to the short field lines (30\unit{Mm}, dark blue at the upper left) with the strongest Ohmic heating.

Taken all together, this supports the interpretation that the magnetic energy input into a field line is responsible for the coronal Ohmic heating and that a threshold of about 200\unit{G} in the footpoint flux density is a prerequisite to providing a sufficient net vertical Poynting flux.
A sufficiently strong flux density also allows for a field line that reaches into the corona, instead of closing at low height to one of the many neighboring opposite polarities of less than 200\unit{G}.

\section{Conclusions\label{S:conclusions}}
We found the low density to be a prerequisite for a strong rise of the Ohmic heating per particle and hence the coronal loop-top temperature (see \fig{heating_profile}).
This is interesting because a lower pressure could instead be the result of strong heating that expands the plasma adiabatically.
However, we see equally strong volumetric Ohmic heating in large parts of our model corona and only where the density is low does the temperature rise later.
This supports the idea that in the real corona the volumetric heating is also distributed in space more evenly than the observations of very confined regions with EUV-bright loops would suggest.

The RTV temperature scaling law fits roughly to our hottest and strongest heated coronal field lines, while this scaling law does not hold for less heated or cooler field lines.
Looking only at the field lines originating at the center of the coronal loops, our model data deviates from the RTV temperature scaling law by 25\unit{\%}, see symbols in \fig{HR_ohm_total_Temp_max}.
This indicates that the RTV scaling law is not per se applicable to any coronal field line, but is roughly consistent with EUV-emissive coronal loops that this scaling law was originally fitted to.

Regarding the discussion whether loops are heated from their footpoints or along the loop, we can say that in our model loops, the volumetric heating is of course stratified following the expansion of the magnetic field (see \fig{heating_profile}).
However, when looking at the heating rates per particle (\fig{heating_particle_profile}), we see that it is actually strongest around 9\unit{Mm} height.
Therefore, speaking of ``footpoint heating'' (in absolute volumetric values) is physically correct, while at the same time it is irrelevant for basic properties of the coronal loops, instead it would be better to better speak of ``loop-leg heating'', because even the term ``loop-top heating'' would be incorrect here \citep[cf.][]{Priest+al:1998}.

\vspace{20pt}

This finding is perfectly consistent with the field-line braiding mechanism, because the photospheric horizontal motions will have the strongest impact through currents generated by deformation or displacement of a magnetic field line where the field is still rather vertical.
The resulting heating per particle is strongest where the density is low enough.
Both conditions are optimally fulfilled for coronal loops in their ``legs'', which we refer to as the lower coronal part of a loop at heights around 5--10\unit{Mm}.

Fundamental in this model is that the heating is due to a slow magnetic energy dissipation process, which would be described best as steady Ohmic heating due to quasi-static magnetic field changes induced by photospheric advection leading to a slow reconfiguration with simultaneously generated electric currents.


\begin{acknowledgements}
This work was supported by the International Max-Planck Research School (IMPRS) on Solar System Physics and was partially funded by the Max-Planck-Princeton Center for Plasma Physics (MPPC).
The results of this research have been achieved using the PRACE Research Infrastructure resource \emph{Curie} based in France at TGCC, as well as \emph{JuRoPA} hosted by the J{\"u}lich Supercomputing Centre in Germany.
Preparatory work has been executed at the Kiepenheuer-Institut f{\"u}r Sonnenphysik in Freiburg, as well as on the bwGRiD facility located at the Universit{\"a}t Freiburg, Germany.
We thank Suguru Kamio for his help finding active region observations.
Hinode is a Japanese mission developed, launched, and operated by ISAS/JAXA, in partnership with NAOJ, NASA, and STFC (UK). Additional operational support is provided by ESA and NSC (Norway).
\end{acknowledgements}

\bibliography{Literatur}
\bibliographystyle{aa}

\end{document}

%% file: Journal.inc
\renewcommand*\aap{A\&A}

\renewcommand*\ao{Appl.~Opt.}

\renewcommand*\apjl{ApJL}

\renewcommand*\apjs{ApJS}